\documentstyle[aps,prl,epsfig,floats]{revtex}

\begin{document}
\draft
\wideabs{

\title{Statistical mechanics of money}

\author{Adrian Dr\u{a}gulescu and Victor M.\ Yakovenko\cite{Yakovenko}}

\address{Department of Physics, University of Maryland, College Park,
MD 20742-4111, USA}

\date{\bf cond-mat/0001432, final version 4, August 4, 2000, to be
published in Eur.~Phys.~J.~B}

\maketitle

\begin{abstract}%
In a closed economic system, money is conserved.  Thus, by analogy
with energy, the equilibrium probability distribution of money must
follow the exponential Boltzmann-Gibbs law characterized by an
effective temperature equal to the average amount of money per
economic agent.  We demonstrate how the Boltzmann-Gibbs distribution
emerges in computer simulations of economic models.  Then we consider
a thermal machine, in which the difference of temperatures allows one
to extract a monetary profit.  We also discuss the role of debt, and
models with broken time-reversal symmetry for which the
Boltzmann-Gibbs law does not hold.  The instantaneous distribution of
money among the agents of a system should not be confused with the
distribution of wealth.  The latter also includes material wealth,
which is not conserved, and thus may have a different (e.\ g.\
power-law) distribution.
\end{abstract}

}

\section{Introduction}

The application of statistical physics methods to economics promises
fresh insights into problems traditionally not associated with physics
(see, for example, the recent review and book \cite{Farmer}).  Both
statistical mechanics and economics study big ensembles: collections
of atoms or economic agents, respectively.  The fundamental law of
equilibrium statistical mechanics is the Boltzmann-Gibbs law, which
states that the probability distribution of energy $\varepsilon$ is
$P(\varepsilon)=Ce^{-\varepsilon/T}$, where $T$ is the temperature,
and $C$ is a normalizing constant \cite{StatPhys}.  The main
ingredient that is essential for the textbook derivation of the
Boltzmann-Gibbs law \cite{StatPhys} is the conservation of energy
\cite{Tsallis}.  Thus, one may generalize that any conserved quantity
in a big statistical system should have an exponential probability
distribution in equilibrium.

An example of such an unconventional Boltzmann-Gibbs law is the
probability distribution of forces experienced by the beads in a
cylinder pressed with an external force \cite{Nagel}.  Because the
system is at rest, the total force along the cylinder axis experienced
by each layer of granules is constant and is randomly distributed
among the individual beads.  Thus the conditions are satisfied for the
applicability of the Boltzmann-Gibbs law to the force, rather than
energy, and it was indeed found experimentally \cite{Nagel}.

We claim that, in a closed economic system, the total amount of money
is conserved.  Thus the equilibrium probability distribution of money
$P(m)$ should follow the Boltzmann-Gibbs law $P(m)=Ce^{-m/T}$.  Here
$m$ is money, and $T$ is an effective temperature equal to the average
amount of money per economic agent.  The conservation law of money
\cite{Shubik} reflects their fundamental property that, unlike
material wealth, money (more precisely the fiat, ``paper'' money) is
not allowed to be manufactured by regular economic agents, but can
only be transferred between agents.  Our approach here is very similar
to that of Ispolatov {\it et al.}  \cite{Redner}.  However, they
considered only models with broken time-reversal symmetry, for which
the Boltzmann-Gibbs law typically does not hold.  The role of
time-reversal symmetry and deviations from the Boltzmann-Gibbs law are
discussed in detail in Sec.\ \ref{Non-Gibbs}.

It is tempting to identify the money distribution $P(m)$ with the
distribution of wealth \cite{Redner}.  However, money is only one part
of wealth, the other part being material wealth.  Material products
have no conservation law: They can be manufactured, destroyed,
consumed, etc.  Moreover, the monetary value of a material product
(the price) is not constant.  The same applies to stocks, which
economics textbooks explicitly exclude from the definition of money
\cite{McConnell}.  So, in general, we do not expect the
Boltzmann-Gibbs law for the distribution of wealth.  Some authors
believe that wealth is distributed according to a power law
(Pareto-Zipf), which originates from a multiplicative random process
\cite{Montroll}.  Such a process may reflect, among other things, the
fluctuations of prices needed to evaluate the monetary value of
material wealth.

\section{Boltzmann-Gibbs distribution}

Let us consider a system of many economic agents $N\gg1$, which may be
individuals or corporations.  In this paper, we only consider the case
where their number is constant.  Each agent $i$ has some money $m_i$
and may exchange it with other agents.  It is implied that money is
used for some economic activity, such as buying or selling material
products; however, we are not interested in that aspect.  As in Ref.\
\cite{Redner}, for us the only result of interaction between agents
$i$ and $j$ is that some money $\Delta m$ changes hands:
$[m_i,m_j]\to[m_i',m_j']=[m_i-\Delta m,m_j+\Delta m]$.  Notice that
the total amount of money is conserved in each transaction:
$m_i+m_j=m_i'+m_j'$.  This local conservation law of money
\cite{Shubik} is analogous to the conservation of energy in collisions
between atoms.  We assume that the economic system is closed, i.\ e.\
there is no external flux of money, thus the total amount of money $M$
in the system is conserved.  Also, in the first part of the paper, we
do not permit any debt, so each agent's money must be non-negative:
$m_i\geq0$.  A similar condition applies to the kinetic energy of
atoms: $\varepsilon_i\geq0$.

Let us introduce the probability distribution function of money
$P(m)$, which is defined so that the number of agents with money
between $m$ and $m+dm$ is equal to $NP(m)\,dm$.  We are interested in
the stationary distribution $P(m)$ corresponding to the state of
thermodynamic equilibrium.  In this state, an individual agent's money
$m_i$ strongly fluctuates, but the overall probability distribution
$P(m)$ does not change.

The equilibrium distribution function $P(m)$ can be derived in the
same manner as the equilibrium distribution function of energy
$P(\varepsilon)$ in physics \cite{StatPhys}.  Let us divide the system
into two subsystems 1 and 2.  Taking into account that money is
conserved and additive: $m=m_1+m_2$, whereas the probability is
multiplicative: $P=P_1P_2$, we conclude that
$P(m_1+m_2)=P(m_1)P(m_2)$.  The solution of this equation is
$P(m)=Ce^{-m/T}$; thus the equilibrium probability distribution of
money has the Boltzmann-Gibbs form.  From the normalization conditions
$\int_0^\infty P(m)\,dm=1$ and $\int_0^\infty m\,P(m)\,dm=M/N$, we
find that $C=1/T$ and $T=M/N$.  Thus, the effective temperature $T$ is
the average amount of money per agent.

The Boltzmann-Gibbs distribution can be also obtained by maximizing
the entropy of money distribution $S=-\int_0^\infty dm\,P(m)\ln P(m)$
under the constraint of money conservation \cite{StatPhys}.  Following
original Boltzmann's argument, let us divide the money axis $0\leq
m\leq\infty$ into small bins of size $dm$ and number the bins
consecutively with the index $b=1,2,\dots\;$ Let us denote the number
of agents in a bin $b$ as $N_b$, the total number being
$N=\sum_{b=1}^{\infty}N_b$.  The agents in the bin $b$ have money
$m_b$, and the total money is $M=\sum_{b=1}^{\infty}m_bN_b$.  The
probability of realization of a certain set of occupation numbers
$\{N_b\}$ is proportional to the number of ways $N$ agents can be
distributed among the bins preserving the set $\{N_b\}$.  This number
is $N!/N_1!N_2!\ldots\;$ The logarithm of probability is entropy $\ln
N!-\sum_{b=1}^{\infty}\ln N_b!$.  When the numbers $N_b$ are big and
Stirling's formula $\ln N!\approx N\ln N$ applies, the entropy per
agent is $S=(N\ln N-\sum_{b=1}^{\infty}N_b\ln
N_b)/N=-\sum_{b=1}^{\infty}P_b\ln P_b$, where $P_b=N_b/N$ is the
probability that an agent has money $m_b$.  Using the method of
Lagrange multipliers to maximize the entropy $S$ with respect to the
occupation numbers $\{N_b\}$ with the constraints on the total money
$M$ and the total number of agents $N$ generates the Boltzmann-Gibbs
distribution for $P(m)$ \cite{StatPhys}.

\section{Computer simulations}

\begin{figure}
\centerline{\epsfig{file=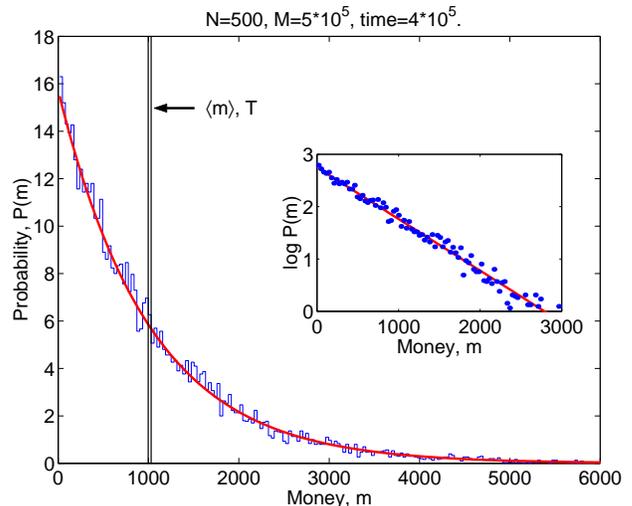,width=0.95\linewidth}}
\caption{Histogram and points: stationary probability distribution of
money $P(m)$. Solid curves: fits to the Boltzmann-Gibbs law
$P(m)\propto\exp(-m/T)$.  Vertical lines: the initial distribution of
money.}
\label{fig:model3}
\end{figure}

To check that these general arguments indeed work, we performed
several computer simulations.  Initially, all agents are given the
same amount of money: $P(m)=\delta(m-M/N)$, which is shown in Fig.\
\ref{fig:model3} as the double vertical line.  One pair of agents at a
time is chosen randomly, then one of the agents is randomly picked to
be the ``winner'' (the other agent becomes the ``loser''), and the
amount $\Delta m\geq0$ is transferred from the loser to the winner.
If the loser does not have enough money to pay ($m_i<\Delta m$), then
the transaction does not take place, and we proceed to another pair of
agents.  Thus, the agents are not permitted to have negative money.
This boundary condition is crucial in establishing the stationary
distribution.  As the agents exchange money, the initial
delta-function distribution first spread symmetrically.  Then, the
probability density starts to pile up at the impenetrable boundary
$m=0$.  The distribution becomes asymmetric (skewed) and ultimately
reaches the stationary exponential shape shown in Fig.\
\ref{fig:model3}.  We used several trading rules in the simulations:
the exchange of a small constant amount $\Delta m=1$, the exchange of
a random fraction $0\le\nu\le1$ of the average money of the pair:
$\Delta m=\nu(m_i+m_j)/2$, and the exchange of a random fraction $\nu$
of the average money in the system: $\Delta m=\nu\,M/N$.  Figures in
the paper mostly show simulations for the third rule; however, the
final stationary distribution was found to be the same for all rules.

\begin{figure}
\centerline{\epsfig{file=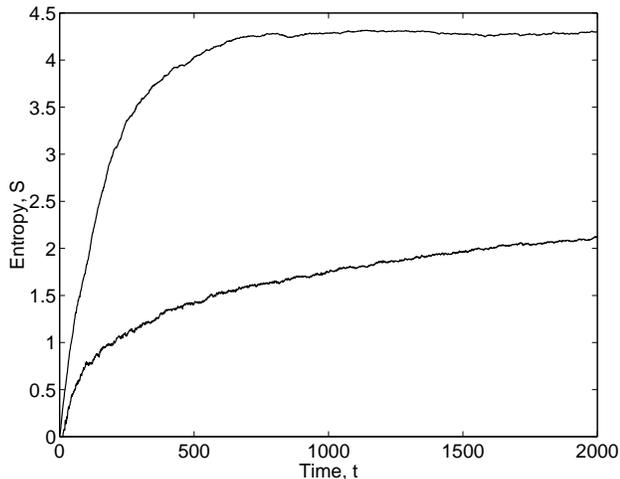,width=0.95\linewidth}}
\caption{Time evolution of entropy.  Top curve: for the exchange of a
random fraction $\nu$ of the average money in the system: $\Delta
m=\nu\,M/N$.  Bottom curve: for the exchange of a small constant
amount $\Delta m=1$.  The time scale for the bottom curve is 500 times
greater than indicated, so it actually ends at the time $10^6$.}
\label{fig:entropy}
\end{figure}

In the process of evolution, the entropy $S$ increases in time and
saturates at the maximal value for the Boltzmann-Gibbs distribution.
This is illustrated by the top curve in Fig.\ \ref{fig:entropy}
computed for the third rule of exchange.  The bottom curve in Fig.\
\ref{fig:entropy} shows the time evolution of entropy for the first
rule of exchange.  The time scale for this curve is 500 times greater
than for the top curve, so the bottom curve actually ends at the time
$10^6$.  The plot shows that, for the first rule of exchange, mixing
is much slower than for the third one.  Nevertheless, even for the
first rule, the system also eventually reaches the Boltzmann-Gibbs
state of maximal entropy, albeit over a time much longer than shown in
Fig.\ \ref{fig:entropy}.

One might argue that the pairwise exchange of money may correspond to
a medieval market, but not to a modern economy.  In order to make the
model somewhat more realistic, we introduce firms.  One agent at a
time becomes a ``firm''.  The firm borrows capital $K$ from another
agent and returns it with an interest $rK$, hires $L$ agents and pays
them wages $W$, manufactures $Q$ items of a product and sell it to $Q$
agents at a price $R$.  All of these agents are randomly selected.
The firm receives the profit $F=RQ-LW-rK$.  The net result is a
many-body exchange of money that still satisfies the conservation law.

Parameters of the model are selected following the procedure described
in economics textbooks.  The aggregate demand-supply curve for the
product is taken to be $R(Q)=V/Q^\eta$, where $Q$ is the quantity
people would buy at a price $R$, and $\eta=0.5$ and $V=100$ are
constants.  The production function of the firm has the conventional
Cobb-Douglas form: $Q(L,K)=L^\beta K^{1-\beta}$, where $\beta=0.8$ is
a constant.  In our simulation, we set $W=10$.  By maximizing firm's
profit $F$ with respect to $K$ and $L$, we find the values of the
other parameters: $L=20$, $Q=10$, $R=32$, and $F=68$.

However, the actual values of the parameters do not matter.  Our
computer simulations show that the stationary probability distribution
of money in this model always has the universal Boltzmann-Gibbs form
independent of the model parameters.

\section{Thermal machine}

As explained in Introduction, the money distribution $P(m)$ should not
be confused with the distribution of wealth.  We believe that $P(m)$
should be interpreted as the instantaneous distribution of purchasing
power in the system.  Indeed, to make a purchase, one needs money.
Material wealth normally is not used directly for a purchase.  It
needs to be sold first to be converted into money.

Let us consider an outside monopolistic vendor selling a product (say,
cars) to the system of agents at a price $p$.  Suppose that a certain
small fraction $f$ of the agents needs to buy the product at a given
time, and each agent who has enough money to afford the price will buy
one item.  The fraction $f$ is assumed to be sufficiently small, so
that the purchase does not perturb the whole system significantly.  At
the same time, the absolute number of agents in this group is assumed
to be big enough to make the group statistically representative and
characterized by the Boltzmann-Gibbs distribution of money.  The
agents in this group continue to exchange money with the rest of the
system, which acts as a thermal bath.  The demand for the product is
constantly renewed, because products purchased in the past expire
after a certain time.  In this situation, the vendor can sell the
product persistently, thus creating a small steady leakage of money
from the system to the vendor.

What price $p$ would maximize the vendor's income?  To answer this
question, it is convenient to introduce the cumulative distribution of
purchasing power ${\cal N}(m)=N\int_m^\infty P(m')\,dm'=Ne^{-m/T}$,
which gives the number of agents whose money is greater than $m$.  The
vendor's income is $fp{\cal N}(p)$.  It is maximal when $p=T$, i.\ e.\
the optimal price is equal to the temperature of the system.  This
conclusion also follows from the simple dimensional argument that
temperature is the only money scale in the problem.  At the price
$p=T$ that maximizes the vendor's income, only the fraction ${\cal
N}(T)/N=e^{-1}=0.37$ of the agents can afford to buy the product.

Now let us consider two disconnected economic systems, one with the
temperature $T_1$ and another with $T_2$: $T_1>T_2$.  A vendor can buy
a product in the latter system at its equilibrium price $T_2$ and sell
it in the former system at the price $T_1$, thus extracting the
speculative profit $T_1-T_2$, as in a thermal machine.  This example
suggests that speculative profit is possible only when the system as a
whole is out of equilibrium.  As money is transferred from the high-
to the low-temperature system, their temperatures become closer and
eventually equal.  After that, no speculative profit is possible,
which would correspond to the ``thermal death'' of the economy.  This
example brings to mind economic relations between developed and
developing countries, with manufacturing in the poor (low-temperature)
countries for export to the rich (high-temperature) ones.

\section{Models with debt}

\begin{figure}
\centerline{\epsfig{file=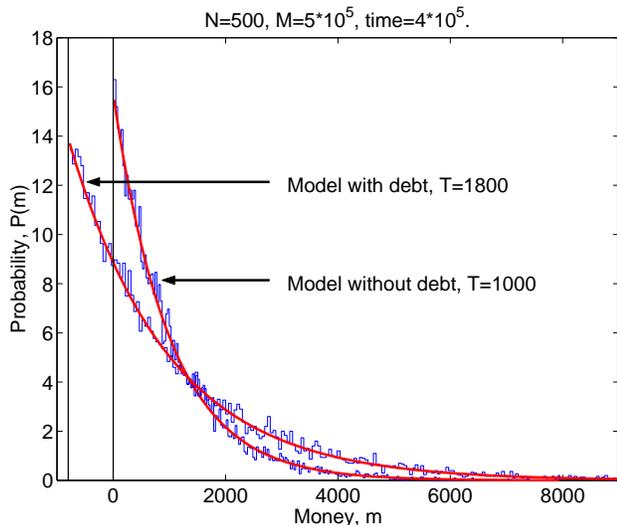,width=0.95\linewidth}}
\caption{Histograms: stationary distributions of money with and
without debt.  Solid curves: fits to the Boltzmann-Gibbs laws with
temperatures $T=1800$ and $T=1000$.}
\label{fig:modeldebt}
\end{figure}

Now let us discuss what happens if the agents are permitted to go into
debt.  Debt can be viewed as negative money.  Now when a loser does
not have enough money to pay, he can borrow the required amount from a
reservoir, and his balance becomes negative.  The conservation law is
not violated: The sum of the winner's positive money and loser's
negative money remains constant.  When an agent with a negative
balance receives money as a winner, she uses this money to repay the
debt until her balance becomes positive.  We assume for simplicity
that the reservoir charges no interest for the lent money.  However,
because it is not sensible to permit unlimited debt, we put a limit
$m_d$ on the maximal debt of an agent: $ m_i>-m_d$.  This new boundary
condition $P(m<-m_d)=0$ replaces the old boundary condition
$P(m<0)=0$.  The result of a computer simulation with $m_d=800$ is
shown in Fig.\ \ref{fig:modeldebt} together with the curve for
$m_d=0$.  $P(m)$ is again given by the Boltzmann-Gibbs law, but now
with the higher temperature $T=M/N+m_d$, because the normalization
conditions need to be maintained including the population with
negative money: $\int_{-m_d}^\infty P(m)\,dm=1$ and
$\int_{-m_d}^\infty m\,P(m)\,dm=M/N$.  The higher temperature makes
the money distribution broader, which means that debt increases
inequality between agents \cite{T<0}.

Imposing a sharp cutoff at $m_d$ may be not quite realistic.  In
practice, the cutoff may be extended over some range depending on the
exact bankruptcy rules.  Over this range, the Boltzmann-Gibbs
distribution would be smeared out.  So we expect to see the
Boltzmann-Gibbs law only sufficiently far from the cutoff region.
Similarly, in experiment \cite{Nagel}, some deviations from the
exponential law were observed near the lower boundary of the
distribution.  Also, at the high end of the distributions, the number
of events becomes small and statistics poor, so the Boltzmann-Gibbs
law loses applicability.  Thus, we expect the Boltzmann-Gibbs law to
hold only for the intermediate range of money not too close either to
the lower boundary or to the very high end.  However, this range is
the most relevant, because it covers the great majority of population.

Lending creates equal amounts of positive (asset) and negative
(liability) money \cite{Shubik,McConnell}.  When economics textbooks
describe how ``banks create money'' or ``debt creates money''
\cite{McConnell}, they do not count the negative liabilities as money,
and thus their money is not conserved.  In our operational definition
of money, we include all financial instruments with fixed
denomination, such as currency, IOUs, and bonds, but not material
wealth or stocks, and we count both assets and liabilities.  With this
definition, money is conserved, and we expect to see the
Boltzmann-Gibbs distribution in equilibrium.  Unfortunately, because
this definition differs from economists' definitions of money (M1, M2,
M3, etc.\ \cite{McConnell}), it is not easy to find the appropriate
statistics.  Of course, money can be also emitted by a central bank or
government.  This is analogous to an external influx of energy into a
physical system.  However, if this process is sufficiently slow, the
economic system may be able to maintain quasi-equilibrium,
characterized by a slowly changing temperature.

We performed a simulation of a model with one bank and many agents.
The agents keep their money in accounts on which the bank pays
interest.  The agents may borrow money from the bank, for which they
must pay interest in monthly installments.  If they cannot make the
required payments, they may be declared bankrupt, which relieves them
from the debt, but the liability is transferred to the bank.  In this
way, the conservation of money is maintained.  The model is too
elaborate to describe it in full detail here.  We found that,
depending on the parameters of the model, either the agents constantly
lose money to the bank, which steadily reduces the agents'
temperature, or the bank constantly loses money, which drives down its
own negative balance and steadily increases the agents' temperature.

\section{Boltzmann equation}

The Boltzmann-Gibbs distribution can be also derived from the
Boltzmann equation \cite{Kinetics}, which describes the time evolution
of the distribution function $P(m)$ due to pairwise interactions: \FL
\begin{eqnarray} 
  &&\frac{dP(m)}{dt}=\int\!\int\{
    -w_{[m,m']\to[m-\Delta,m'+\Delta]}P(m)P(m')
\label{Boltzmann}  \\
  &&+w_{[m-\Delta,m'+\Delta]\to[m,m']}
  P(m-\Delta)P(m'+\Delta)\}\,dm'\,d\Delta.  \nonumber
\end{eqnarray} 
Here $w_{[m,m']\to[m-\Delta,m'+\Delta]}$ is the rate of transferring
money $\Delta$ from an agent with money $m$ to an agent with money
$m'$.  If a model has time-reversal symmetry, then the transition rate
of a direct process is the same as the transition rate of the reversed
process, thus the $w$-factors in the first and second lines of Eq.\
(\ref{Boltzmann}) are equal.  In this case, the Boltzmann-Gibbs
distribution $P(m)=C\exp(-m/T)$ nullifies the right-hand side of Eq.\
(\ref{Boltzmann}); thus this distribution is stationary: $dP(m)/dt=0$
\cite{Kinetics}.

\section{Non-Boltzmann-Gibbs distributions}
\label{Non-Gibbs}

\begin{figure}
\centerline{\epsfig{file=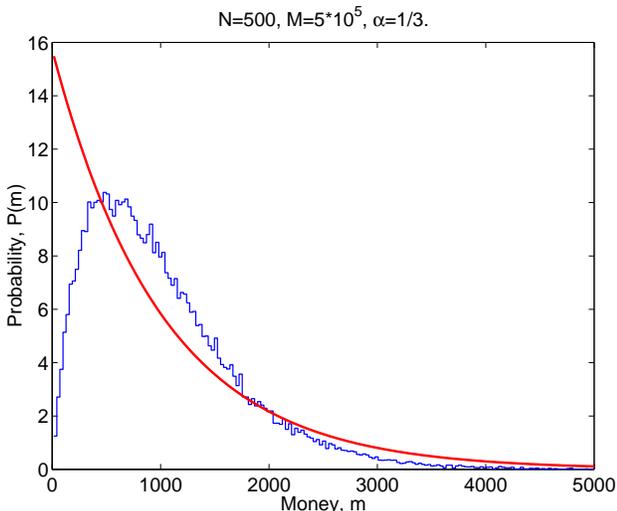,width=0.95\linewidth}}
\caption{Histogram: stationary probability distribution of money in
the multiplicative random exchange model studied in Ref.\
\protect\cite{Redner}.  Solid curve: the Boltzmann-Gibbs law.}
\label{fig:rednerplot}
\end{figure}

However, if time-reversal symmetry is broken, the two transition rates
$w$ in Eq.\ (\ref{Boltzmann}) may be different, and the system may
have a non-Boltzmann-Gibbs stationary distribution or no stationary
distribution at all.  Examples of such kind were studied in Ref.\
\cite{Redner}.  One model was called the multiplicative random
exchange.  In this model, a randomly selected loser $i$ loses a fixed
fraction $\alpha$ of his money $m_i$ to a randomly selected winner
$j$: $[m_i,m_j]\to[(1-\alpha)m_i\:,\:m_j+\alpha m_i]$.  If we try to
reverse this process and appoint the winner $j$ to become a loser, the
system does not return to the original configuration $[m_i,m_j]$:
$[(1-\alpha)m_i\:,\:m_j+\alpha m_i]\to[(1-\alpha)m_i+\alpha(m_j+\alpha
m_i)\:,\:(1-\alpha)(m_j+\alpha m_i)]$.  Except for $\alpha=1/2$, the
exponential distribution function is not a stationary solution of the
Boltzmann equation derived for this model in Ref.\ \cite{Redner}.
Instead, the stationary distribution has the shape shown in Fig.\
\ref{fig:rednerplot} for $\alpha=1/3$, which we reproduced in our
numerical simulations.  It still has an exponential tail end at the
high end, but drops to zero at the low end for $\alpha<1/2$.  Another
example of similar kind was studied in Ref.\ \cite{Chakraborti}, which
appeared after the first version of our paper was posted as
cond-mat/0001432 on January 30, 2000.  In that model, the agents save
a fraction $\lambda$ of their money and exchange a random fraction
$\epsilon$ of their total remaining money: $[m_i, m_j] \to [\lambda
m_i + \epsilon(1-\lambda)(m_i+m_j)\:,\: \lambda m_j +
(1-\epsilon)(1-\lambda)(m_i+m_j)]$.  This exchange also does not
return to the original configuration after being reversed.  The
stationary probability distribution was found in Ref.\
\cite{Chakraborti} to be nonexponential for $\lambda\neq0$ with a
shape qualitatively similar to the one shown in Fig.\
\ref{fig:rednerplot}.

\begin{figure}
\centerline{\epsfig{file=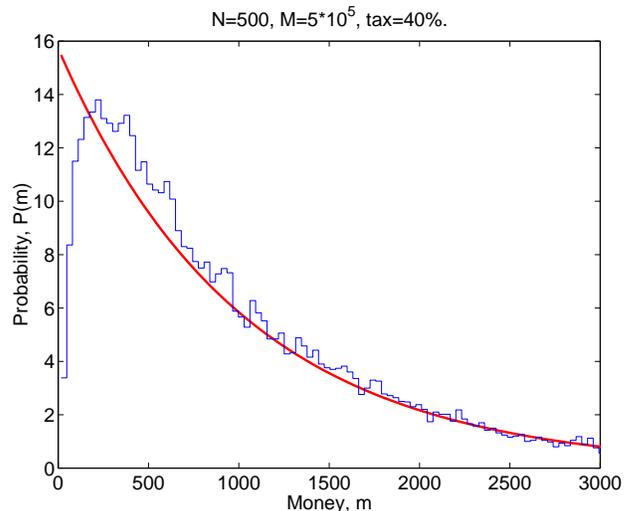,width=0.95\linewidth}}
\caption{Histogram: stationary probability distribution of money in
the model with taxes and subsidies.  Solid curve: the Boltzmann-Gibbs
law.}
\label{fig:govmod}
\end{figure}

Another interesting example of a non-Boltzmann-Gibbs distribution
occurs in a model with taxes and subsidies.  Suppose a special agent
(``government'') collects a fraction (``tax'') of every transaction in
the system.  The collected money is then equally divided between all
agents of the system, so that each agent receives the subsidy $\delta
m$ with the frequency $1/\tau_s$.  Assuming that $\delta m$ is small
and approximating the collision integral with a relaxation time
$\tau_r$ \cite{Kinetics}, we obtain the following Boltzmann equation
\begin{equation} 
  \frac{\partial P(m)}{\partial t}+\frac{\delta m}{\tau_s}\,
  \frac{\partial P(m)}{\partial m}=-\frac{P(m)-\tilde{P}(m)}{\tau_r},
\label{tax}
\end{equation} 
where $\tilde{P}(m)$ is the equilibrium Boltzmann-Gibbs function.  The
second term in the left-hand side of Eq.\ (\ref{tax}) is analogous to
the force applied to electrons in a metal by an external electric
field \cite{Kinetics}.  The approximate stationary solution of Eq.\
(\ref{tax}) is the displaced Boltzmann-Gibbs one:
$P(m)=\tilde{P}(m-(\tau_r/\tau_s)\,\delta m)$.  The displacement of
the equilibrium distribution $\tilde{P}(m)$ by
$(\tau_r/\tau_s)\,\delta m$ would leave an empty gap near $m=0$.  This
gap is filled by interpolating between zero population at $m=0$ and
the displaced distribution.  The curve obtained in a computer
simulation of this model (Fig.\ \ref{fig:govmod}) qualitatively agrees
with this expectation.  The low-money population is suppressed,
because the government, acting as an external force, ``pumps out''
that population and pushes the system out of thermodynamic
equilibrium.  We found that the entropy of the stationary state in the
model with taxes and subsidies is few percents lower than without.

These examples show that the Boltzmann-Gibbs distribution is not fully
universal, meaning that it does not hold for just any model of
exchange that conserves money.  Nevertheless, it is universal in a
limited sense: For a broad class of models that have time-reversal
symmetry, the stationary distribution is exponential and does not
depend on the details of a model.  Conversely, when time-reversal
symmetry is broken, the distribution may depend on model details.  The
difference between these two classes of models may be rather subtle.
For example, let us change the multiplicative random exchange from a
fixed fraction of loser's money to a fixed fraction of the total money
of winner and loser.  This modification retains the multiplicative
idea that the amount exchanged is proportional to the amount involved,
but restores time-reversal symmetry and the Boltzmann-Gibbs
distribution.  In the model with $\Delta m=1$ discussed in the next
Section, the difference between time-reversible and time-irreversible
formulations amounts to the difference between impenetrable and
absorbing boundary conditions at $m=0$.  Unlike in physics, in economy
there is no fundamental requirement that interactions have
time-reversal symmetry.  However, in the absence of detailed knowledge
of real microscopic dynamics of economic exchange, the semiuniversal
Boltzmann-Gibbs distribution appears to be a natural starting point.

Moreover, deviations from the Boltzmann-Gibbs law may occur only if
the transition rates $w$ in Eq.\ (\ref{Boltzmann}) explicitly depend
on the agents money $m$ or $m'$ in an asymmetric manner.  In another
simulation, we randomly preselected winners and losers for every pair
of agents $(i,j)$.  In this case, money flows along directed links
between the agents: $i\!\to\!j\!\to\!k$, and time-reversal symmetry is
strongly broken.  This model is closer to the real economy, in which,
for example, one typically receives money from an employer and pays it
to a grocer, but rarely the reverse.  Nevertheless, we still found the
Boltzmann-Gibbs distribution of money in this model, because the
transition rates $w$ do not explicitly depend on $m$ and $m'$.

\section{Nonlinear Boltzmann equation vs. linear master equation}

For the model where agents randomly exchange the constant amount
$\Delta m=1$, the Boltzmann equation is: \FL
\begin{eqnarray} 
  \frac{dP_m}{dt}&=&P_{m+1}\sum\limits_{n=0}^\infty P_n
     +P_{m-1}\sum\limits_{n=1}^\infty P_n 
\nonumber \\
  &&{}-P_m\sum\limits_{n=0}^\infty P_n 
 -P_m\sum\limits_{n=1}^\infty P_n 
\label{D0} \\
  &=&(P_{m+1}+P_{m-1}-2P_m)+P_0(P_m-P_{m-1}),
\label{D1}
\end{eqnarray} 
where $P_m\equiv P(m)$ and we have used $\sum_{m=0}^\infty P_m=1$.
The first, diffusion term in Eq.\ (\ref{D1}) is responsible for
broadening of the initial delta-function distribution.  The second
term, proportional to $P_0$, is essential for the Boltzmann-Gibbs
distribution $P_m=e^{-m/T}(1-e^{-1/T})$ to be a stationary solution of
Eq.\ (\ref{D1}).  In a similar model studied in Ref.\ \cite{Redner},
the second term was omitted on the assumption that agents who lost all
money are eliminated: $P_0=0$.  In that case, the total number of
agents is not conserved, and the system never reaches any stationary
distribution.  Time-reversal symmetry is violated, since transitions
into the state $m=0$ are permitted, but not out of this state.

If we treat $P_0$ as a constant, Eq.\ (\ref{D1}) looks like a linear
Fokker-Planck equation \cite{Kinetics} for $P_m$, with the first term
describing diffusion and the second term an external force
proportional to $P_0$.  Similar equations were studied in Ref.\
\cite{Montroll}.  Eq.\ (\ref{D1}) can be also rewritten as
\begin{equation}
\frac{dP_m}{dt}=P_{m+1}-(2-P_0)P_m+(1-P_0)P_{m-1}.
\label{Fokker-Planck}
\end{equation}
The coefficient $(1-P_0)$ in front of $P_{m-1}$ represents the rate of
increasing money by $\Delta m=1$, and the coefficient 1 in front of
$P_{m+1}$ represents the rate of decreasing money by $\Delta m=-1$.
Since $P_0>0$, the former is smaller than the latter, which results in
the stationary Boltzmann-Gibbs distributions $P_m=(1-P_0)^m$.  An
equation similar to Eq.\ (\ref{Fokker-Planck}) describes a Markov
chain studied for strategic market games in Ref.\ \cite{Shubik-2}.
Naturally, the stationary probability distribution of wealth in that
model was found to be exponential \cite{Shubik-2}.

Even though Eqs.\ (\ref{D1}) and (\ref{Fokker-Planck}) look like
linear equations, nevertheless the Boltzmann equation
(\ref{Boltzmann}) and (\ref{D0}) is a profoundly nonlinear equation.
It contains the product of two probability distribution functions $P$
in the right-hand side, because two agents are involved in money
exchange.  Most studies of wealth distribution \cite{Montroll} have
the fundamental flaw that they use a single-particle approach.  They
assume that the wealth of an agent may change just by itself and write
a linear master equation for the probability distribution.  Because
only one particle is considered, this approach cannot adequately
incorporate conservation of money.  In reality, an agent can change
money only by interacting with another agent, thus the problem
requires a two-particle probability distribution function.  Using
Boltzmann's molecular chaos hypothesis, the two-particle function is
factorized into a product of two single-particle distributions
functions, which results in the nonlinear Boltzmann equation.
Conservation of money is adequately incorporated in this two-particle
approach, and the universality of the exponential Boltzmann-Gibbs
distribution is transparent.

\section{Conclusions}

Everywhere in the paper we assumed some randomness in the exchange of
money.  The results of our paper would apply the best to the
probability distribution of money in a closed community of gamblers.
In more traditional economic studies, the agents exchange money not
randomly, but following deterministic strategies, such as maximization
of utility functions \cite{Shubik,Bak}.  The concept of equilibrium in
these studies is similar to mechanical equilibrium in physics, which
is achieved by minimizing energy or maximizing utility.  However, for
big ensembles, statistical equilibrium is a more relevant concept.
When many heterogeneous agents deterministically interact and spend
various amounts of money from very little to very big, the money
exchange is effectively random.  In the future, we would like to
uncover the Boltzmann-Gibbs distribution of money in a simulation of a
big ensemble of economic agents following realistic deterministic
strategies with money conservation taken into account.  That would be
the economics analog of molecular dynamics simulations in physics.
While atoms collide following fully deterministic equations of motion,
their energy exchange is effectively random due to complexity of the
system and results in the Boltzmann-Gibbs law.

We do not claim that the real economy is in equilibrium.  (Most of the
physical world around us is not in true equilibrium either.)
Nevertheless, the concept of statistical equilibrium is a very useful
reference point for studying nonequilibrium phenomena.

The authors are grateful to M.\ Gubrud for helpful discussion and
proofreading of an earlier version of the manuscript.

{\it Note added:} After the paper had been submitted for publication,
we have learned about the book by Aoki \cite{Aoki}, who applied many
ideas of statistical physics to economics, albeit not specifically to
the distribution of money.

%\vspace{-1.5\baselineskip}


\begin{references}

%\vspace{-4.5\baselineskip}

\bibitem[*]{Yakovenko} E-mail: yakovenk@physics.umd.edu \\
Web: http://www2.physics.umd.edu/\~{}yakovenk

\bibitem{Farmer} J. D. Farmer, Computing in Science and Engineering
    {\bf 1}, \#6, p. 26 (1999); R. N. Mantegna and H. E. Stanley, {\it An
    Introduction to Econophysics} (Cambridge University Press,
    Cambridge, 2000).
  
\bibitem{StatPhys} G. H. Wannier, {\it Statistical Physics} (Dover,
    New York, 1987).

\bibitem{Tsallis} It is also implied that the system is extensive.  We
   study only extensive models, so the nonextensive generalization of
   entropy by C. Tsallis, J. Stat. Phys. {\bf 52}, 479 (1988);
   Braz. J. Phys. {\bf 29}, 1 (1999) is not relevant for our
   consideration.

\bibitem{Nagel} D. M. Mueth, H. M. Jaeger, and S. R. Nagel,
   Phys. Rev. E {\bf 57}, 3164 (1998); M. L. Nguyen and
   S. N. Coppersmith, Phys. Rev. E {\bf 59}, 5870 (1999).
 
\bibitem{Shubik} M. Shubik, in {\it The Economy as an Evolving Complex
   System II}, edited by W. B. Arthur, S. N. Durlauf, and D. A. Lane
   (Addison-Wesley, Reading, 1997) p. 263.
  
\bibitem{Redner} S. Ispolatov, P. L. Krapivsky, and S. Redner, Eur.
   Phys. J. B {\bf 2}, 267 (1998).
  
\bibitem{McConnell} C. R. McConnell and S. L. Brue, {\it Economics:
   Principles, Problems, and Policies} (McGraw-Hill, New York, 1996).
  
\bibitem{Montroll} E. W. Montroll and M. F. Shlesinger,
   Proc. Natl. Acad. Sci. USA {\bf 79}, 3380 (1982); O. Malcai,
   O. Biham, and S. Solomon, Phys. Rev. E {\bf 60}, 1299 (1999);
   D. Sornette and R. Cont, J. Phys. I (France) {\bf 7}, 431 (1997);
   J.-P. Bouchaud and M. M\'ezard, cond-mat/0002374.

\bibitem{T<0} In general, temperature is completely determined by the
   average money per agent, $\langle m\rangle=M/N$, and the boundary
   conditions.  Suppose the agents are required to have no less money
   than $m_1$ and no more than $m_2$: $m_1\leq m\leq m_2$.  In this
   case, the normalization conditions $\int_{m_1}^{m_2}P(m)\,dm=1$ and
   $\int_{m_1}^{m_2} m\,P(m)\,dm=\langle m\rangle$ with
   $P(m)=C\,e^{-m/T}$ give the following equation for $T$:
\begin{equation}
  \coth\left(\frac{\Delta m}{T}\right)-\frac{T}{\Delta m}=
  \frac{\overline{m}-\langle m\rangle}{\Delta m},
\label{T}
\end{equation}
   where $\overline{m}=(m_1+m_2)/2$ and $\Delta m=(m_2-m_1)/2$.  It
   follows from Eq.\ (\ref{T}) that the temperature is positive when
   $\overline{m}>\langle m\rangle$, negative when
   $\overline{m}<\langle m\rangle$, and infinite ($P(m)=\rm const$)
   when $\overline{m}=\langle m\rangle$.  In particular, if agents'
   money are bounded from above, but not from below: $-\infty\leq
   m\leq m_2$, the temperature is negative.  That means inverted
   Boltzmann-Gibbs distribution with more reach agents than poor.

\bibitem{Kinetics} E. M. Lifshitz and L. P. Pitaevski\u{\i}, {\it
   Physical Kinetics} (Pergamon Press, New York, 1993).
 
\bibitem{Chakraborti} A. Chakraborti and B. K. Chakrabarty,
   cond-mat/0004256, to be published in Eur. Phys. J. B.

\bibitem{Shubik-2} Martin Shubik, {\it The Theory of Money and Financial
   Institutions}, vol. 2 (The MIT Press, Cambridge, 1999), p. 192.

\bibitem{Bak} P. Bak, S. F. N{\o}rrelykke, and M.  Shubik,
   Phys. Rev. E {\bf 60}, 2528 (1999).
  
\bibitem{Aoki} Masanao Aoki, {\it New Approaches to Macroeconomic
   Modeling} (Cambridge University Press, Cambridge, 1996).

\end{references}
\end{document}